\documentclass[prl,showpacs,amsmath,amssymb,twocolumn]{revtex4-1}
\usepackage{graphicx}

\begin{document}

\title{Random fields at a nonequilibrium phase transition}

\author{Hatem Barghathi and  Thomas Vojta}
\affiliation{Department of Physics, Missouri University of Science $\&$ Technology, Rolla, MO 65409}

\date{\today}

\begin{abstract}
We study nonequilibrium phase transitions in the presence of disorder that
locally breaks the symmetry between two equivalent macroscopic states.
In low-dimensional equilibrium systems, such ``random-field'' disorder is known to have dramatic
effects: It prevents spontaneous symmetry breaking and  completely destroys the
phase transition. In contrast, we show that the phase transition of the
one-dimensional generalized contact process persists in the presence of random field
disorder. The ultraslow dynamics in the
symmetry-broken phase is described by a Sinai walk of the domain
walls between two different absorbing states. We discuss the generality and limitations
of our theory, and we illustrate our results by large-scale Monte-Carlo simulations.
\end{abstract}


\maketitle


Impurities, defects, and other types of quenched disorder can have drastic effects on the
long-time and large-distance behavior of many-particle systems. For example, disorder can
modify the universality class of a critical point \cite{HarrisLubensky74,GrinsteinLuther76},
change a phase transition from first order to continuous
\cite{ImryWortis79,HuiBerker89,AizenmanWehr89}, or smear a sharp transition over an interval
of the tuning parameter \cite{Vojta03a,*Vojta03b,*Vojta04}.
Particularly strong effects arise from disorder that \emph{locally} breaks the symmetry
between two equivalent macroscopic states while preserving the symmetry globally (in the
statistical sense). As this type of disorder corresponds to a random external field in a magnetic
system, it is usually called random-field disorder.
Recently, a beautiful example of a random-field magnet was discovered
in {LiHo$_x$Y$_{1-x}$F$_4$} \cite{TGKSF06,SBBGAR07,Schechter08}.
Random-field disorder naturally occurs when the order parameter
breaks a \emph{real-space} symmetry such in as nematic liquid crystals in porous media
\cite{MCBB94} and stripe states in high-temperature superconductors \cite{CDFK06}.

Imry and Ma \cite{ImryMa75} discussed random-field effects at equilibrium
phase transitions based on an appealing heuristic argument. Consider a uniform domain of linear size $L$
in $d$ space dimensions.
The free energy gain due to aligning this domain
with the (average) local random field behaves as $L^{d/2}$
while the domain wall energy is of the order of
$L^{d-1}$ \footnote{This holds for discrete symmetry. For continuous symmetry
the surface energy behaves as $L^{d-2}$ resulting in a marginal dimension of 4}.
For $d<2$, the system thus gains free energy by forming finite-size domains that
align with the random field. In contrast, for $d>2$, the uniform state is preferred.
Building on this work, Aizenman and Wehr \cite{AizenmanWehr89} proved rigorously
that random-field disorder prevents spontaneous symmetry breaking in all dimensions
$d\le2$ for Ising symmetry and $d\le4$ for continuous symmetry. Thus, random fields destroy
an equilibrium phase transition in sufficiently low dimensions.

In nature, thermal equilibrium is rather the exception than the rule. Although equilibrium is an
excellent approximation for some systems, many others are far from equilibrium and show
qualitatively different behaviors. In recent years, phase transitions between
different nonequilibrium states have attracted considerable attention. Examples can
be found in population dynamics, chemical reactions, growing surfaces, granular flow
as well as traffic jams
\cite{SchmittmannZia95,MarroDickman99,Hinrichsen00,Odor04,TauberHowardVollmayrLee05}.
It is therefore important to study random-field effects at
such nonequilibrium phase transitions. Are these transitions destroyed by random fields
just like equilibrium transitions?

\begin{figure}
\includegraphics[width=8.7cm]{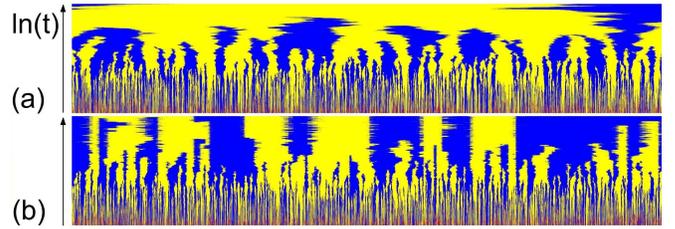}
\caption{(Color online) Time evolution of the generalized contact process in the inactive phase:
(a) without ($\mu=5/6$) and (b) with random-field disorder ($\mu_h=1, \mu_l=2/3$). 
$I_1$ and $I_2$ are shown in yellow and blue (light and dark grey). Active sites
between the domains are marked in red (middle grey).
The difference between the diffusive domain wall motion (a) and the much slower Sinai walk (b) is
clearly visible (part of a system of $10^5$ sites for times up to $10^8$).  }
\label{fig:evo}
\end{figure}

In this Letter, we address this question for a prominent class of nonequilibrium phase
transitions, \emph{viz.}, absorbing state transitions separating active, fluctuating states
from inactive, absorbing states where fluctuations cease entirely. We develop a heuristic argument
showing that
random-field disorder which locally favors one of two equivalent absorbing states over the other
does \emph{not} prevent global spontaneous symmetry breaking in any dimension. The random fields
thus do \emph{not} destroy the nonequilibrium transition.
In the symmetry-broken phase, the relevant degrees of freedom are domain walls between
different absorbing states. Their long-time dynamics is given by a Sinai walk
\cite{Solomon75,*KestenKozlovSpitzer75,*Sinai82}
leading to an ultraslow approach to the absorbing state during which the density
of domain walls decays as $\ln^{-2}(t)$ with time $t$ (see Fig.\ \ref{fig:evo}).
We also study the behavior right at the
critical point where we find even slower dynamics.

In the remainder of the Letter, we sketch the derivation of the results; and we support them
by Monte-Carlo simulations. For definiteness, we first consider the generalized contact
process with two absorbing states \cite{Hinrichsen97} in one dimension. We later argue
that our heuristic argument applies to an entire class of absorbing state transitions.


The (simple) contact process \cite{HarrisTE74} is a prototypical model featuring
an absorbing state transition. Each site of a $d$-dimensional hypercubic lattice
is either in the active (infected) state $A$ or in the inactive (healthy) state $I$.
The time evolution is a continuous-time Markov process with
infected sites healing at a rate $\mu$ while healthy sites become infected at a rate $\lambda m /(2d)$
where $m$ is the number of infected nearest neighbors. The long-time behavior is governed
by the ratio of  $\lambda$ and $\mu$. If $\mu \gg \lambda$, healing dominates over infection,
and all sites will eventually be healthy. The absorbing state without any infected sites is
thus the only steady state. For $\lambda \gg \mu$, the
infection never dies out, leading to an active steady state with  nonzero density of infected sites.
The absorbing and active steady states are separated by a nonequilibrium transition in the
directed percolation (DP) \cite{GrassbergerdelaTorre79} universality class.

Following Hinrichsen \cite{Hinrichsen97}, we generalize the contact process by allowing each site
to be in one of $n+1$ states, the active state $A$ or one of $n$ inactive states $I_k$ ($k=1\ldots n$).
The time evolution of the generalized contact process (GCP) is conveniently defined \cite{Hinrichsen97}
via the transition rates for pairs of nearest-neighbor sites,
\begin{eqnarray}
w(A A \to A I_k) = w(A A \to I_k A) &=& \bar\mu_k/n~,
\label{eq:rate_barmu}\\
w(A I_k \to I_k I_k) = w(I_k A \to I_k I_k) &=& \mu_k~,
\label{eq:rate_mu}\\
w(A I_k \to A A) = w(I_k A \to A A) &=& \lambda~,
\label{eq:rate_lambda}\\
w(I_k I_l \to I_k A) = w(I_k I_l \to A I_l) &=&
\sigma~,
\label{eq:rate_sigma}
\end{eqnarray}
with $k,l=1\ldots n$ and $k \ne l$. All other rates vanish.
The GCP defined by (\ref{eq:rate_barmu}) to (\ref{eq:rate_sigma}) reduces to
the simple contact process if we set $n=1$ and $\bar \mu_k = \mu_k = \mu$
(up to rescaling all rates by the same constant factor
\footnote{The rescaling factor is the number of nearest-neighbor pairs a site belongs to; for a hypercubic lattice it is $2d$.}).
The transition (\ref{eq:rate_sigma}) permits competition between different inactive states as it prevents
different domains from sticking together. Instead, they
can separate, and the domain walls can move.
We now set $\bar\mu_k = \mu_k$ and $\lambda = \sigma =1$ to keep the parameter space manageable
\footnote{According to Ref.\ \cite{LeeVojta10}, the qualitative behavior for
          $\bar \mu_k  \ne \mu_k$ is identical to that for $\bar \mu_k = \mu_k$. Moreover, the precise value
          of $\sigma$ is not important as long as it is nonzero.}.
This also fixes the time unit. Moreover, we focus on
$d=1$ and $n=2$.

The long-time behavior again follows from comparing the infection rate $\lambda$ with the
healing rates $\mu_1$ and $\mu_2$. Consider two equivalent inactive states, $\mu_1 = \mu_2 = \mu$.
For small $\mu$, the system is in the active phase with nonzero density of
infected sites. In this fluctuating phase, the symmetry between the two inactive states $I_1$ and $I_2$
is not broken, i.e., their occupancies are identical. If $\mu$ is increased beyond
$\mu_c^0 \approx 0.628$ \cite{Hinrichsen97,LeeVojta10}, the system
undergoes a nonequilibrium phase transition to one of the two absorbing steady states (either all sites in state $I_1$
or all in state $I_2$). At this transition, the symmetry between
$I_1$ and $I_2$ is \emph{spontaneously} broken. Its critical behavior is therefore not
in the DP universality class but
in the so-called DP2 class which, in $d=1$, coincides with the parity conserving (PC)
class \cite{GrassbergerKrauseTwer84}. If $\mu_1 \ne \mu_2$,
one of the two inactive states dominates for long times, and the critical behavior reverts back to DP.


We introduce quenched (time-independent) disorder by making the healing rates $\mu_k(\mathbf{r})$ at
site $\mathbf{r}$ independent random variables governed by a probability distribution $W(\mu_1,\mu_2)$.
As we are interested in random-field disorder which locally breaks
the symmetry between $I_1$ and $I_2$, we choose $\mu_1(\mathbf{r}) \ne \mu_2(\mathbf{r})$.
Globally, the symmetry is preserved in the statistical sense
implying $W(\mu_1,\mu_2) = W(\mu_2,\mu_1)$. An example is the correlated binary distribution
\begin{equation}
W(\mu_1,\mu_2) = \frac 1 2 \delta(\mu_1 - \mu_h)\delta(\mu_2-\mu_l) + \frac 1 2 \delta(\mu_1 - \mu_l)\delta(\mu_2-\mu_h)
\label{eq:distrib}
\end{equation}
with possible local healing rate values $\mu_h$ or $\mu_l$
\footnote{Other disorder types can lead to  different behaviors
\cite{OdorMenyhard06,*MenyhardOdor07}}.

To address our main question, namely whether the random-field disorder prevents
the spontaneous breaking of the global symmetry between the two inactive states and thus destroys
the nonequilibrium transition, we analyze the large-$\mu$ regime where all healing rates are larger than the
clean critical value $\mu_c^0$. In this regime, almost all sites quickly decay into one of the
two inactive states $I_1$ or $I_2$. The relevant long-time degrees of freedom are domain walls between
$I_1$ and $I_2$ domains. They move via a combination of process (\ref{eq:rate_sigma}) which creates
an active site at the domain wall and process (\ref{eq:rate_mu}) which allows this active site to decay
into either $I_1$ or $I_2$. Because of the disorder, the resulting domain wall hopping rates depend on the site $\mathbf{r}$.
Importantly, the rates for hopping right and left are different because the underlying healing rates
$\mu_1(\mathbf{r})$ and $\mu_2(\mathbf{r})$ are not identical.

The long-time dynamics in the large-$\mu$ regime is thus governed by a random walk of the domain walls.
Due to the local left-right asymmetry, this random walk is not a conventional (diffusive) walk
but a Sinai walk \footnote{Because our model preserves the global
symmetry between $I_1$ and $I_2$, the Sinai walk is unbiased.}.
The typical displacement of a Sinai walker grows as $\ln^2(t/t_0)$
with time $t$ \cite{Solomon75,*KestenKozlovSpitzer75,*Sinai82} ($t_0$ is a microscopic time scale), more slowly than the
well-known $t^{1/2}$ law for a conventional walk (see Fig.\ \ref{fig:evo}).
When two neighboring domain walls meet, they annihilate, replacing
three domains by a single one. Domain walls surviving at time $t$ thus have a typical
distance proportional to $\ln^2(t/t_0)$. The domains grow without limit, and
their density decays as $\ln^{-2}(t/t_0)$. In the long-time limit, the system reaches
a single-domain state, i.e., either all sites are in state $I_1$ or all in $I_2$. This implies that the symmetry
between $I_1$ and $I_2$ is spontaneously broken (which of the two absorbing states the system ends up in
depends on details of the initial conditions and of the stochastic time evolution).
The nonequilibrium transition consequently persists in the presence of random-field disorder.

It is important to contrast the domain wall dynamics in our system with that of a corresponding equilibrium
problem such as the random-field Ising chain (whose low-temperature state consists of domains of up and down
spins
\footnote{In this analogy, we relate the ordered spin-up and spin-down states of the random-field Ising chain to the two
absorbing states of the GCP.}).
The crucial difference is that the inactive states $I_1$ and $I_2$ in our system are absorbing:
Active sites and new domain walls never arise in the interior of a domain. In contrast, inside a uniform domain of
the random-field Ising chain, a spin flip (which creates two new domain walls) can occur anywhere due to a thermal
fluctuation. This mechanism limits the growth of the typical domain size to its equilibrium value dictated
by the Imry-Ma argument \cite{ImryMa75}, and thus prevents spontaneous symmetry breaking.

\begin{figure}[tb]
\includegraphics[width=8.7cm]{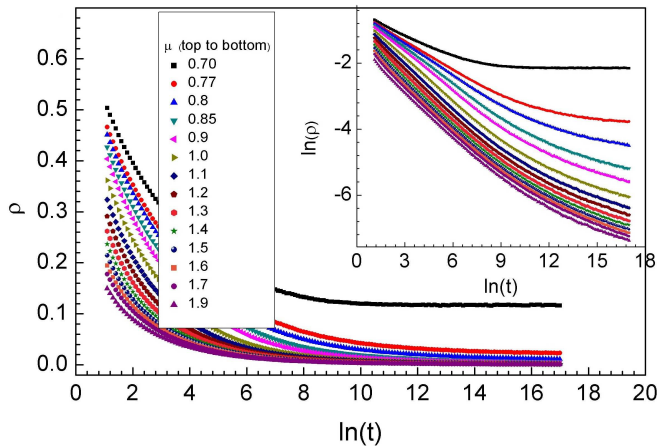}
\caption{(Color online) Density $\rho$ vs. time $t$ for several values of the healing rate $\mu$.
         The data are averages over
         60 to 200 disorder configurations. Inset: The log-log plot shows that the density decay is slower than
         a power law for all $\mu$.}
\label{fig:overview}
\end{figure}

To verify these heuristic arguments and to illustrate the results, we perform
Monte-Carlo simulations \cite{LeeVojta10} of the one-dimensional GCP with random-field disorder.
We use system sizes up to $L=10^5$ and times up to $t=2\times 10^8$.
The random-field disorder is implemented via the
distribution (\ref{eq:distrib}) with $1.5 \mu_l = \mu_h \equiv \mu$.
Our simulations start from a fully active lattice (all sites in state A),
and we monitor the density $\rho$ of active sites as well as the densities $\rho_1$ and $\rho_2$
of sites in the inactive states $I_1$ and $I_2$, respectively.
Figure \ref{fig:overview} presents an overview of the time evolution of the density $\rho$.

\begin{figure}[tb]
\includegraphics[width=8.7cm]{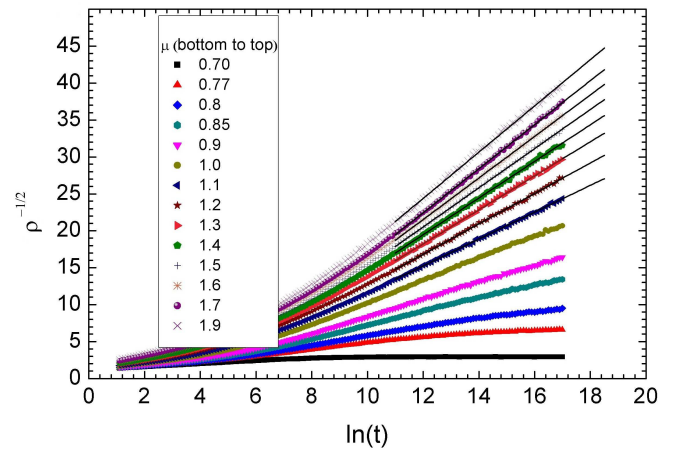}
\caption{(Color online) $\rho^{-1/2}$ vs. $\ln(t)$ for several values of the healing rate $\mu$. The solid straight lines
        are fits to the predicted behavior $\rho \sim \ln^{-2}(t/t_0)$.}
\label{fig:sinai}
\end{figure}

We now focus on the curves with healing rates $\mu \gtrsim 1.0$ for which both $\mu_h=\mu$ and $\mu_l = 2\mu/3$ are larger than the clean critical value $\mu_c^0$.
The inset of Fig.\ \ref{fig:overview} shows that the density continues to decay to the longest times studied
for all these curves. However, the decay is clearly slower than a power law. To compare with our theoretical arguments,
we note that active sites only exist near domain walls in the large-$\mu$ regime. We thus expect the density of
active sites to be proportional to the domain wall density,  yielding $\rho \sim \ln^{-2}(t/t_0)$.
To test this prediction we plot $\rho^{-1/2}$ vs. $\ln(t)$ in Fig.\ \ref{fig:sinai};
in such a graph the expected behavior corresponds to a straight line.
The figure shows that all curves with $\mu >1$ indeed follow the prediction over several
orders of magnitude in time.

\begin{figure}[tb]
\includegraphics[width=8.7cm]{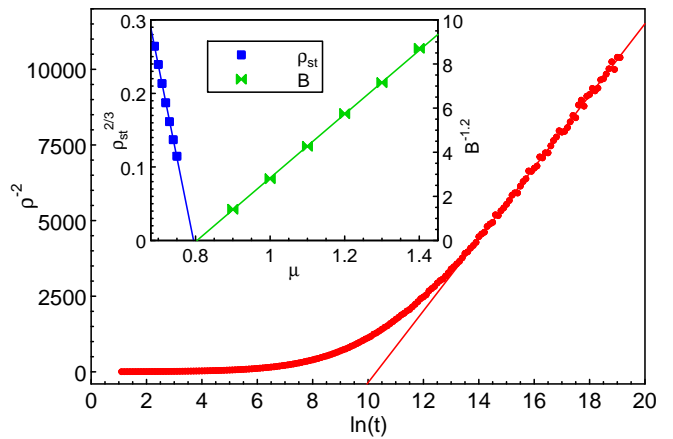}
\caption{(Color online) Density vs. time at the critical healing rate $\mu_c=0.8$,
  plotted as $\rho^{-1/x}$ vs. $\ln(t)$ with $x = 0.5$. The solid line is a fit
  to $\rho(t) \sim \ln^{-x}(t/t_0)$.
  Inset: Identifying $\mu_c$ from the stationary density $\rho_{st}$
  in the active phase and the prefactor of the $\rho =B \ln^{-2}(t/t_0)$ decay in the inactive phase.}
\label{fig:critical}
\end{figure}

In addition to the inactive phase, we also study the critical point.
To identify the critical healing rate $\mu_c$, we extrapolate to zero
both the stationary density $\rho_{st} = \lim_{t\to \infty} \rho(t)$ in the active phase and
the inverse prefactor of the $\ln^{-2}(t/t_0)$ decay in the inactive phase.
This  yields $\mu_c \approx 0.80$ (see inset of Fig.\ \ref{fig:critical}).
At this healing rate, the density decay is clearly slower than the $\ln^{-2}(t/t_0)$ law
governing the inactive phase. This extremely slow decay and the uncertainty in $\mu_c$
prevent us from determining the functional form of the critical $\rho(t)$ curve unambiguously.
If we assume a time dependence of the type $\rho(t) \sim \ln^{-x}(t/t_0)$ we find a value of $x \approx 0.5$.
Moreover, from the dependence of the stationary density on the healing rate, $\rho_{st} \sim (\mu_c-\mu)^\beta$,
we obtain $\beta \approx 1.5$. The values of $x$ and $\beta$ should be considered rough estimates.
An accurate determination of the critical behavior of the GCP
 with random-field disorder requires a significantly larger numerical effort and remains a task
for the future.


In summary, we have shown that the nonequilibrium phase transition
of the one-dimensional GCP  survives in the presence of random-field disorder,
in contrast to one-dimensional equilibrium
transitions that are destroyed by random fields. In the concluding paragraphs, we discuss
the generality and limitations of our results.

The crucial difference between random-field effects in equilibrium systems
such as the random-field Ising chain and in the GCP is the
absorbing character of the inactive states $I_1$ and $I_2$ in the latter. The interior
of an $I_1$ or $I_2$ domain is ``dead'' as no active sites and no new domain walls
can ever arise there. In contrast, in an equilibrium system, pairs of new domain walls can appear
in the interior of a uniform domain via a thermal fluctuation. This limits the growth of the
typical domain size to the Imry-Ma equilibrium size and thus destroys the equilibrium transition
(in sufficiently low dimensions).
We expect our results to hold for all nonequilibrium phase transition at which the random-field disorder
locally breaks the symmetry between two \emph{absorbing} states. Other nonequilibrium transitions may behave differently.
For example, our theory does \emph{not} apply if the random fields break the symmetry between two active states.

In the symmetry-broken inactive phase, the dynamics of the GCP with random-field disorder
is ultraslow. It is governed by the Sinai random walk of domain walls between the two inactive states.
 This leads to a logarithmic time decay of the densities of both domain walls
and active sites.
Note that the Sinai coarsening dynamics has been studied in detail in the equilibrium random-field Ising chain
\cite{FisherLeDoussalMonthus98,*FisherLeDoussalMonthus01} where it applies to a transient time
regime before the domains reach the Imry-Ma equilibrium size.

Although our explicit results are for one dimension, we expect our main conclusion to hold in higher dimensions, too.
In the interior
of a uniform domain of an absorbing state, new active site (and new domain walls) cannot arise in any dimension. Moreover,
the Imry-Ma mechanism by which the random fields destroy an equilibrium transition becomes less effective in
higher dimensions. Indeed, Pigolotti and Cencini \cite{PigolottiCencini10} report spontaneous symmetry breaking
in a model of two species competing in a two-dimensional landscape with local habitat preferences.
To further study this question, we plan to introduce random fields into
our simulations of the two-dimensional GCP \cite{LeeVojta11}.

Finally, we turn to experiments. Although clear-cut realizations of absorbing state transitions were lacking
for a long time \cite{Hinrichsen00b}, beautiful examples were recently found in
turbulent liquid crystals \cite{TKCS07}, driven suspensions \cite{CCGP08,FFGP11}, and superconducting
vortices \cite{OkumaTsugawaMotohashi11}. As they are far from equilibrium, biological systems
are promising candidates for observing nonequilibrium transitions. A transition in
the DP2 universality class (as studied here) occurs in a model of competing bacteria
strains \cite{KorolevNelson11} which accurately describes experiments in colony biofilms \cite{KXNF11}.
Random-field disorder could be realized in such experiments by environments that locally favor one
strain over the other.


We thank M. Mu\~noz and G. Odor for helpful discussions.
This work has been supported by the NSF under Grant Nos. DMR-0906566
and DMR-1205803.

\bibliographystyle{apsrev4-1}
\bibliography{../../00Bibtex/rareregions}

\begin{thebibliography}{46}%
\makeatletter
\providecommand \@ifxundefined [1]{%
 \@ifx{#1\undefined}
}%
\providecommand \@ifnum [1]{%
 \ifnum #1\expandafter \@firstoftwo
 \else \expandafter \@secondoftwo
 \fi
}%
\providecommand \@ifx [1]{%
 \ifx #1\expandafter \@firstoftwo
 \else \expandafter \@secondoftwo
 \fi
}%
\providecommand \natexlab [1]{#1}%
\providecommand \enquote  [1]{``#1''}%
\providecommand \bibnamefont  [1]{#1}%
\providecommand \bibfnamefont [1]{#1}%
\providecommand \citenamefont [1]{#1}%
\providecommand \href@noop [0]{\@secondoftwo}%
\providecommand \href [0]{\begingroup \@sanitize@url \@href}%
\providecommand \@href[1]{\@@startlink{#1}\@@href}%
\providecommand \@@href[1]{\endgroup#1\@@endlink}%
\providecommand \@sanitize@url [0]{\catcode `\\12\catcode `\$12\catcode
  `\&12\catcode `\#12\catcode `\^12\catcode `\_12\catcode `\%12\relax}%
\providecommand \@@startlink[1]{}%
\providecommand \@@endlink[0]{}%
\providecommand \url  [0]{\begingroup\@sanitize@url \@url }%
\providecommand \@url [1]{\endgroup\@href {#1}{\urlprefix }}%
\providecommand \urlprefix  [0]{URL }%
\providecommand \Eprint [0]{\href }%
\providecommand \doibase [0]{http://dx.doi.org/}%
\providecommand \selectlanguage [0]{\@gobble}%
\providecommand \bibinfo  [0]{\@secondoftwo}%
\providecommand \bibfield  [0]{\@secondoftwo}%
\providecommand \translation [1]{[#1]}%
\providecommand \BibitemOpen [0]{}%
\providecommand \bibitemStop [0]{}%
\providecommand \bibitemNoStop [0]{.\EOS\space}%
\providecommand \EOS [0]{\spacefactor3000\relax}%
\providecommand \BibitemShut  [1]{\csname bibitem#1\endcsname}%
\let\auto@bib@innerbib\@empty
\bibitem [{\citenamefont {Harris}\ and\ \citenamefont
  {Lubensky}(1974)}]{HarrisLubensky74}%
  \BibitemOpen
  \bibfield  {author} {\bibinfo {author} {\bibfnamefont {A.~B.}\ \bibnamefont
  {Harris}}\ and\ \bibinfo {author} {\bibfnamefont {T.~C.}\ \bibnamefont
  {Lubensky}},\ }\href@noop {} {\bibfield  {journal} {\bibinfo  {journal}
  {Phys. Rev. Lett.}\ }\textbf {\bibinfo {volume} {33}},\ \bibinfo {pages}
  {1540} (\bibinfo {year} {1974})}\BibitemShut {NoStop}%
\bibitem [{\citenamefont {Grinstein}\ and\ \citenamefont
  {Luther}(1976)}]{GrinsteinLuther76}%
  \BibitemOpen
  \bibfield  {author} {\bibinfo {author} {\bibfnamefont {G.}~\bibnamefont
  {Grinstein}}\ and\ \bibinfo {author} {\bibfnamefont {A.}~\bibnamefont
  {Luther}},\ }\href@noop {} {\bibfield  {journal} {\bibinfo  {journal} {Phys.
  Rev. B}\ }\textbf {\bibinfo {volume} {13}},\ \bibinfo {pages} {1329}
  (\bibinfo {year} {1976})}\BibitemShut {NoStop}%
\bibitem [{\citenamefont {Imry}\ and\ \citenamefont
  {Wortis}(1979)}]{ImryWortis79}%
  \BibitemOpen
  \bibfield  {author} {\bibinfo {author} {\bibfnamefont {Y.}~\bibnamefont
  {Imry}}\ and\ \bibinfo {author} {\bibfnamefont {M.}~\bibnamefont {Wortis}},\
  }\href@noop {} {\bibfield  {journal} {\bibinfo  {journal} {Phys. Rev. B}\
  }\textbf {\bibinfo {volume} {19}},\ \bibinfo {pages} {3580} (\bibinfo {year}
  {1979})}\BibitemShut {NoStop}%
\bibitem [{\citenamefont {Hui}\ and\ \citenamefont
  {Berker}(1989)}]{HuiBerker89}%
  \BibitemOpen
  \bibfield  {author} {\bibinfo {author} {\bibfnamefont {K.}~\bibnamefont
  {Hui}}\ and\ \bibinfo {author} {\bibfnamefont {A.~N.}\ \bibnamefont
  {Berker}},\ }\href@noop {} {\bibfield  {journal} {\bibinfo  {journal} {Phys.
  Rev. Lett.}\ }\textbf {\bibinfo {volume} {62}},\ \bibinfo {pages} {2507}
  (\bibinfo {year} {1989})}\BibitemShut {NoStop}%
\bibitem [{\citenamefont {Aizenman}\ and\ \citenamefont
  {Wehr}(1989)}]{AizenmanWehr89}%
  \BibitemOpen
  \bibfield  {author} {\bibinfo {author} {\bibfnamefont {M.}~\bibnamefont
  {Aizenman}}\ and\ \bibinfo {author} {\bibfnamefont {J.}~\bibnamefont
  {Wehr}},\ }\href@noop {} {\bibfield  {journal} {\bibinfo  {journal} {Phys.
  Rev. Lett.}\ }\textbf {\bibinfo {volume} {62}},\ \bibinfo {pages} {2503}
  (\bibinfo {year} {1989})}\BibitemShut {NoStop}%
\bibitem [{\citenamefont {Vojta}(2003{\natexlab{a}})}]{Vojta03a}%
  \BibitemOpen
  \bibfield  {author} {\bibinfo {author} {\bibfnamefont {T.}~\bibnamefont
  {Vojta}},\ }\href@noop {} {\bibfield  {journal} {\bibinfo  {journal} {Phys.
  Rev. Lett.}\ }\textbf {\bibinfo {volume} {90}},\ \bibinfo {pages} {107202}
  (\bibinfo {year} {2003}{\natexlab{a}})}\BibitemShut {NoStop}%
\bibitem [{\citenamefont {Vojta}(2003{\natexlab{b}})}]{Vojta03b}%
  \BibitemOpen
  \bibfield  {author} {\bibinfo {author} {\bibfnamefont {T.}~\bibnamefont
  {Vojta}},\ }\href@noop {} {\bibfield  {journal} {\bibinfo  {journal} {J.
  Phys. A}\ }\textbf {\bibinfo {volume} {36}},\ \bibinfo {pages} {10921}
  (\bibinfo {year} {2003}{\natexlab{b}})}\BibitemShut {NoStop}%
\bibitem [{\citenamefont {Vojta}(2004)}]{Vojta04}%
  \BibitemOpen
  \bibfield  {author} {\bibinfo {author} {\bibfnamefont {T.}~\bibnamefont
  {Vojta}},\ }\href@noop {} {\bibfield  {journal} {\bibinfo  {journal} {Phys.
  Rev. E}\ }\textbf {\bibinfo {volume} {70}},\ \bibinfo {pages} {026108}
  (\bibinfo {year} {2004})}\BibitemShut {NoStop}%
\bibitem [{\citenamefont {Tabei}\ \emph {et~al.}(2006)\citenamefont {Tabei},
  \citenamefont {Gingras}, \citenamefont {Kao}, \citenamefont {Stasiak},\ and\
  \citenamefont {Fortin}}]{TGKSF06}%
  \BibitemOpen
  \bibfield  {author} {\bibinfo {author} {\bibfnamefont {S.~M.~A.}\
  \bibnamefont {Tabei}}, \bibinfo {author} {\bibfnamefont {M.~J.~P.}\
  \bibnamefont {Gingras}}, \bibinfo {author} {\bibfnamefont {Y.-J.}\
  \bibnamefont {Kao}}, \bibinfo {author} {\bibfnamefont {P.}~\bibnamefont
  {Stasiak}}, \ and\ \bibinfo {author} {\bibfnamefont {J.-Y.}\ \bibnamefont
  {Fortin}},\ }\href@noop {} {\bibfield  {journal} {\bibinfo  {journal} {Phys.
  Rev. Lett.}\ }\textbf {\bibinfo {volume} {97}},\ \bibinfo {eid} {237203}
  (\bibinfo {year} {2006})}\BibitemShut {NoStop}%
\bibitem [{\citenamefont {Silevitch}\ \emph {et~al.}(2007)\citenamefont
  {Silevitch}, \citenamefont {Bitko}, \citenamefont {Brooke}, \citenamefont
  {Ghosh}, \citenamefont {Aeppli},\ and\ \citenamefont {Rosenbaum}}]{SBBGAR07}%
  \BibitemOpen
  \bibfield  {author} {\bibinfo {author} {\bibfnamefont {D.~M.}\ \bibnamefont
  {Silevitch}}, \bibinfo {author} {\bibfnamefont {D.}~\bibnamefont {Bitko}},
  \bibinfo {author} {\bibfnamefont {J.}~\bibnamefont {Brooke}}, \bibinfo
  {author} {\bibfnamefont {S.}~\bibnamefont {Ghosh}}, \bibinfo {author}
  {\bibfnamefont {G.}~\bibnamefont {Aeppli}}, \ and\ \bibinfo {author}
  {\bibfnamefont {T.~F.}\ \bibnamefont {Rosenbaum}},\ }\href@noop {} {\bibfield
   {journal} {\bibinfo  {journal} {Nature}\ }\textbf {\bibinfo {volume}
  {448}},\ \bibinfo {pages} {567} (\bibinfo {year} {2007})}\BibitemShut
  {NoStop}%
\bibitem [{\citenamefont {Schechter}(2008)}]{Schechter08}%
  \BibitemOpen
  \bibfield  {author} {\bibinfo {author} {\bibfnamefont {M.}~\bibnamefont
  {Schechter}},\ }\href@noop {} {\bibfield  {journal} {\bibinfo  {journal}
  {Phys. Rev. B}\ }\textbf {\bibinfo {volume} {77}},\ \bibinfo {eid} {020401}
  (\bibinfo {year} {2008})}\BibitemShut {NoStop}%
\bibitem [{\citenamefont {Maritan}\ \emph {et~al.}(1994)\citenamefont
  {Maritan}, \citenamefont {Cieplak}, \citenamefont {Bellini},\ and\
  \citenamefont {Banavar}}]{MCBB94}%
  \BibitemOpen
  \bibfield  {author} {\bibinfo {author} {\bibfnamefont {A.}~\bibnamefont
  {Maritan}}, \bibinfo {author} {\bibfnamefont {M.}~\bibnamefont {Cieplak}},
  \bibinfo {author} {\bibfnamefont {T.}~\bibnamefont {Bellini}}, \ and\
  \bibinfo {author} {\bibfnamefont {J.~R.}\ \bibnamefont {Banavar}},\ }\href
  {\doibase 10.1103/PhysRevLett.72.4113} {\bibfield  {journal} {\bibinfo
  {journal} {Phys. Rev. Lett.}\ }\textbf {\bibinfo {volume} {72}},\ \bibinfo
  {pages} {4113} (\bibinfo {year} {1994})}\BibitemShut {NoStop}%
\bibitem [{\citenamefont {Carlson}\ \emph {et~al.}(2006)\citenamefont
  {Carlson}, \citenamefont {Dahmen}, \citenamefont {Fradkin},\ and\
  \citenamefont {Kivelson}}]{CDFK06}%
  \BibitemOpen
  \bibfield  {author} {\bibinfo {author} {\bibfnamefont {E.~W.}\ \bibnamefont
  {Carlson}}, \bibinfo {author} {\bibfnamefont {K.~A.}\ \bibnamefont {Dahmen}},
  \bibinfo {author} {\bibfnamefont {E.}~\bibnamefont {Fradkin}}, \ and\
  \bibinfo {author} {\bibfnamefont {S.~A.}\ \bibnamefont {Kivelson}},\ }\href
  {\doibase 10.1103/PhysRevLett.96.097003} {\bibfield  {journal} {\bibinfo
  {journal} {Phys. Rev. Lett.}\ }\textbf {\bibinfo {volume} {96}},\ \bibinfo
  {pages} {097003} (\bibinfo {year} {2006})}\BibitemShut {NoStop}%
\bibitem [{\citenamefont {Imry}\ and\ \citenamefont {Ma}(1975)}]{ImryMa75}%
  \BibitemOpen
  \bibfield  {author} {\bibinfo {author} {\bibfnamefont {Y.}~\bibnamefont
  {Imry}}\ and\ \bibinfo {author} {\bibfnamefont {S.-k.}\ \bibnamefont {Ma}},\
  }\href {\doibase 10.1103/PhysRevLett.35.1399} {\bibfield  {journal} {\bibinfo
   {journal} {Phys. Rev. Lett.}\ }\textbf {\bibinfo {volume} {35}},\ \bibinfo
  {pages} {1399} (\bibinfo {year} {1975})}\BibitemShut {NoStop}%
\bibitem [{Note1()}]{Note1}%
  \BibitemOpen
  \bibinfo {note} {This holds for discrete symmetry. For continuous symmetry
  the surface energy behaves as $L^{d-2}$ resulting in a marginal dimension of
  4}\BibitemShut {NoStop}%
\bibitem [{\citenamefont {Schmittmann}\ and\ \citenamefont
  {Zia}(1995)}]{SchmittmannZia95}%
  \BibitemOpen
  \bibfield  {author} {\bibinfo {author} {\bibfnamefont {B.}~\bibnamefont
  {Schmittmann}}\ and\ \bibinfo {author} {\bibfnamefont {R.~K.~P.}\
  \bibnamefont {Zia}},\ }in\ \href@noop {} {\emph {\bibinfo {booktitle} {Phase
  Transitions and Critical Phenomena}}},\ Vol.~\bibinfo {volume} {17},\
  \bibinfo {editor} {edited by\ \bibinfo {editor} {\bibfnamefont
  {C.}~\bibnamefont {Domb}}\ and\ \bibinfo {editor} {\bibfnamefont {J.~L.}\
  \bibnamefont {Lebowitz}}}\ (\bibinfo  {publisher} {Academic},\ \bibinfo
  {address} {New York},\ \bibinfo {year} {1995})\ p.~\bibinfo {pages}
  {1}\BibitemShut {NoStop}%
\bibitem [{\citenamefont {Marro}\ and\ \citenamefont
  {Dickman}(1999)}]{MarroDickman99}%
  \BibitemOpen
  \bibfield  {author} {\bibinfo {author} {\bibfnamefont {J.}~\bibnamefont
  {Marro}}\ and\ \bibinfo {author} {\bibfnamefont {R.}~\bibnamefont
  {Dickman}},\ }\href@noop {} {\emph {\bibinfo {title} {Nonequilibrium Phase
  Transitions in Lattice Models}}}\ (\bibinfo  {publisher} {Cambridge
  University Press},\ \bibinfo {address} {Cambridge},\ \bibinfo {year}
  {1999})\BibitemShut {NoStop}%
\bibitem [{\citenamefont {Hinrichsen}(2000{\natexlab{a}})}]{Hinrichsen00}%
  \BibitemOpen
  \bibfield  {author} {\bibinfo {author} {\bibfnamefont {H.}~\bibnamefont
  {Hinrichsen}},\ }\href@noop {} {\bibfield  {journal} {\bibinfo  {journal}
  {Adv. Phys.}\ }\textbf {\bibinfo {volume} {49}},\ \bibinfo {pages} {815}
  (\bibinfo {year} {2000}{\natexlab{a}})}\BibitemShut {NoStop}%
\bibitem [{\citenamefont {Odor}(2004)}]{Odor04}%
  \BibitemOpen
  \bibfield  {author} {\bibinfo {author} {\bibfnamefont {G.}~\bibnamefont
  {Odor}},\ }\href@noop {} {\bibfield  {journal} {\bibinfo  {journal} {Rev.
  Mod. Phys.}\ }\textbf {\bibinfo {volume} {76}},\ \bibinfo {pages} {663}
  (\bibinfo {year} {2004})}\BibitemShut {NoStop}%
\bibitem [{\citenamefont {T{\"a}uber}\ \emph {et~al.}(2005)\citenamefont
  {T{\"a}uber}, \citenamefont {Howard},\ and\ \citenamefont
  {Vollmayr-Lee}}]{TauberHowardVollmayrLee05}%
  \BibitemOpen
  \bibfield  {author} {\bibinfo {author} {\bibfnamefont {U.~C.}\ \bibnamefont
  {T{\"a}uber}}, \bibinfo {author} {\bibfnamefont {M.}~\bibnamefont {Howard}},
  \ and\ \bibinfo {author} {\bibfnamefont {B.~P.}\ \bibnamefont
  {Vollmayr-Lee}},\ }\href@noop {} {\bibfield  {journal} {\bibinfo  {journal}
  {J. Phys. A}\ }\textbf {\bibinfo {volume} {38}},\ \bibinfo {pages} {R79}
  (\bibinfo {year} {2005})}\BibitemShut {NoStop}%
\bibitem [{\citenamefont {Solomon}(1975)}]{Solomon75}%
  \BibitemOpen
  \bibfield  {author} {\bibinfo {author} {\bibfnamefont {F.}~\bibnamefont
  {Solomon}},\ }\href@noop {} {\bibfield  {journal} {\bibinfo  {journal} {Ann.
  Prob.}\ }\textbf {\bibinfo {volume} {3}},\ \bibinfo {pages} {1} (\bibinfo
  {year} {1975})}\BibitemShut {NoStop}%
\bibitem [{\citenamefont {Kesten}\ \emph {et~al.}(1975)\citenamefont {Kesten},
  \citenamefont {Kozlov},\ and\ \citenamefont
  {Spitzer}}]{KestenKozlovSpitzer75}%
  \BibitemOpen
  \bibfield  {author} {\bibinfo {author} {\bibfnamefont {H.}~\bibnamefont
  {Kesten}}, \bibinfo {author} {\bibfnamefont {M.}~\bibnamefont {Kozlov}}, \
  and\ \bibinfo {author} {\bibfnamefont {F.}~\bibnamefont {Spitzer}},\
  }\href@noop {} {\bibfield  {journal} {\bibinfo  {journal} {Compositio Math.}\
  }\textbf {\bibinfo {volume} {30}},\ \bibinfo {pages} {145} (\bibinfo {year}
  {1975})}\BibitemShut {NoStop}%
\bibitem [{\citenamefont {Sinai}(1982)}]{Sinai82}%
  \BibitemOpen
  \bibfield  {author} {\bibinfo {author} {\bibfnamefont {Y.~G.}\ \bibnamefont
  {Sinai}},\ }\href@noop {} {\bibfield  {journal} {\bibinfo  {journal} {Theor.
  Probab. Appl.}\ }\textbf {\bibinfo {volume} {27}},\ \bibinfo {pages} {256}
  (\bibinfo {year} {1982})}\BibitemShut {NoStop}%
\bibitem [{\citenamefont {Hinrichsen}(1997)}]{Hinrichsen97}%
  \BibitemOpen
  \bibfield  {author} {\bibinfo {author} {\bibfnamefont {H.}~\bibnamefont
  {Hinrichsen}},\ }\href@noop {} {\bibfield  {journal} {\bibinfo  {journal}
  {Phys. Rev. E}\ }\textbf {\bibinfo {volume} {55}},\ \bibinfo {pages} {219}
  (\bibinfo {year} {1997})}\BibitemShut {NoStop}%
\bibitem [{\citenamefont {Harris}(1974)}]{HarrisTE74}%
  \BibitemOpen
  \bibfield  {author} {\bibinfo {author} {\bibfnamefont {T.~E.}\ \bibnamefont
  {Harris}},\ }\href@noop {} {\bibfield  {journal} {\bibinfo  {journal} {Ann.
  Prob.}\ }\textbf {\bibinfo {volume} {2}},\ \bibinfo {pages} {969} (\bibinfo
  {year} {1974})}\BibitemShut {NoStop}%
\bibitem [{\citenamefont {Grassberger}\ and\ \citenamefont {de~la
  Torre}(1979)}]{GrassbergerdelaTorre79}%
  \BibitemOpen
  \bibfield  {author} {\bibinfo {author} {\bibfnamefont {P.}~\bibnamefont
  {Grassberger}}\ and\ \bibinfo {author} {\bibfnamefont {A.}~\bibnamefont
  {de~la Torre}},\ }\href@noop {} {\bibfield  {journal} {\bibinfo  {journal}
  {Ann. Phys. (NY)}\ }\textbf {\bibinfo {volume} {122}},\ \bibinfo {pages}
  {373} (\bibinfo {year} {1979})}\BibitemShut {NoStop}%
\bibitem [{Note2()}]{Note2}%
  \BibitemOpen
  \bibinfo {note} {The rescaling factor is the number of nearest-neighbor pairs
  a site belongs to; for a hypercubic lattice it is $2d$.}\BibitemShut {Stop}%
\bibitem [{Note3()}]{Note3}%
  \BibitemOpen
  \bibinfo {note} {According to Ref.\ \cite {LeeVojta10}, the qualitative
  behavior for $\protect \mathaccentV {bar}016\mu _k \not =\mu _k$ is identical
  to that for $\protect \mathaccentV {bar}016\mu _k = \mu _k$. Moreover, the
  precise value of $\sigma $ is not important as long as it is
  nonzero.}\BibitemShut {Stop}%
\bibitem [{\citenamefont {Lee}\ and\ \citenamefont {Vojta}(2010)}]{LeeVojta10}%
  \BibitemOpen
  \bibfield  {author} {\bibinfo {author} {\bibfnamefont {M.~Y.}\ \bibnamefont
  {Lee}}\ and\ \bibinfo {author} {\bibfnamefont {T.}~\bibnamefont {Vojta}},\
  }\href@noop {} {\bibfield  {journal} {\bibinfo  {journal} {Phys. Rev. E}\
  }\textbf {\bibinfo {volume} {81}},\ \bibinfo {pages} {061128} (\bibinfo
  {year} {2010})}\BibitemShut {NoStop}%
\bibitem [{\citenamefont {Grassberger}\ \emph {et~al.}(1984)\citenamefont
  {Grassberger}, \citenamefont {Krause},\ and\ \citenamefont {von~der
  Twer}}]{GrassbergerKrauseTwer84}%
  \BibitemOpen
  \bibfield  {author} {\bibinfo {author} {\bibfnamefont {P.}~\bibnamefont
  {Grassberger}}, \bibinfo {author} {\bibfnamefont {F.}~\bibnamefont {Krause}},
  \ and\ \bibinfo {author} {\bibfnamefont {T.}~\bibnamefont {von~der Twer}},\
  }\href@noop {} {\bibfield  {journal} {\bibinfo  {journal} {J. Phys. A}\
  }\textbf {\bibinfo {volume} {17}},\ \bibinfo {pages} {L105} (\bibinfo {year}
  {1984})}\BibitemShut {NoStop}%
\bibitem [{Note4()}]{Note4}%
  \BibitemOpen
  \bibinfo {note} {Other disorder types can lead to different behaviors \cite
  {OdorMenyhard06,*MenyhardOdor07}}\BibitemShut {NoStop}%
\bibitem [{Note5()}]{Note5}%
  \BibitemOpen
  \bibinfo {note} {Because our model preserves the global symmetry between
  $I_1$ and $I_2$, the Sinai walk is unbiased.}\BibitemShut {Stop}%
\bibitem [{Note6()}]{Note6}%
  \BibitemOpen
  \bibinfo {note} {In this analogy, we relate the ordered spin-up and spin-down
  states of the random-field Ising chain to the two absorbing states of the
  GCP.}\BibitemShut {Stop}%
\bibitem [{\citenamefont {Fisher}\ \emph {et~al.}(1998)\citenamefont {Fisher},
  \citenamefont {Le~Doussal},\ and\ \citenamefont
  {Monthus}}]{FisherLeDoussalMonthus98}%
  \BibitemOpen
  \bibfield  {author} {\bibinfo {author} {\bibfnamefont {D.~S.}\ \bibnamefont
  {Fisher}}, \bibinfo {author} {\bibfnamefont {P.}~\bibnamefont {Le~Doussal}},
  \ and\ \bibinfo {author} {\bibfnamefont {C.}~\bibnamefont {Monthus}},\
  }\href@noop {} {\bibfield  {journal} {\bibinfo  {journal} {Phys. Rev. Lett}\
  }\textbf {\bibinfo {volume} {80}},\ \bibinfo {pages} {3539} (\bibinfo {year}
  {1998})}\BibitemShut {NoStop}%
\bibitem [{\citenamefont {Fisher}\ \emph {et~al.}(2001)\citenamefont {Fisher},
  \citenamefont {Le~Doussal},\ and\ \citenamefont
  {Monthus}}]{FisherLeDoussalMonthus01}%
  \BibitemOpen
  \bibfield  {author} {\bibinfo {author} {\bibfnamefont {D.~S.}\ \bibnamefont
  {Fisher}}, \bibinfo {author} {\bibfnamefont {P.}~\bibnamefont {Le~Doussal}},
  \ and\ \bibinfo {author} {\bibfnamefont {C.}~\bibnamefont {Monthus}},\
  }\href@noop {} {\bibfield  {journal} {\bibinfo  {journal} {Phys. Rev. E}\
  }\textbf {\bibinfo {volume} {64}},\ \bibinfo {pages} {066107} (\bibinfo
  {year} {2001})}\BibitemShut {NoStop}%
\bibitem [{\citenamefont {Pigolotti}\ and\ \citenamefont
  {Cencini}(2010)}]{PigolottiCencini10}%
  \BibitemOpen
  \bibfield  {author} {\bibinfo {author} {\bibfnamefont {S.}~\bibnamefont
  {Pigolotti}}\ and\ \bibinfo {author} {\bibfnamefont {M.}~\bibnamefont
  {Cencini}},\ }\href@noop {} {\bibfield  {journal} {\bibinfo  {journal} {J.
  Theor. Biology}\ }\textbf {\bibinfo {volume} {265}},\ \bibinfo {pages} {609}
  (\bibinfo {year} {2010})}\BibitemShut {NoStop}%
\bibitem [{\citenamefont {Lee}\ and\ \citenamefont {Vojta}(2011)}]{LeeVojta11}%
  \BibitemOpen
  \bibfield  {author} {\bibinfo {author} {\bibfnamefont {M.~Y.}\ \bibnamefont
  {Lee}}\ and\ \bibinfo {author} {\bibfnamefont {T.}~\bibnamefont {Vojta}},\
  }\href {\doibase 10.1103/PhysRevE.83.011114} {\bibfield  {journal} {\bibinfo
  {journal} {Phys. Rev. E}\ }\textbf {\bibinfo {volume} {83}},\ \bibinfo
  {pages} {011114} (\bibinfo {year} {2011})}\BibitemShut {NoStop}%
\bibitem [{\citenamefont {Hinrichsen}(2000{\natexlab{b}})}]{Hinrichsen00b}%
  \BibitemOpen
  \bibfield  {author} {\bibinfo {author} {\bibfnamefont {H.}~\bibnamefont
  {Hinrichsen}},\ }\href@noop {} {\bibfield  {journal} {\bibinfo  {journal}
  {Braz. J. Phys.}\ }\textbf {\bibinfo {volume} {30}},\ \bibinfo {pages} {69}
  (\bibinfo {year} {2000}{\natexlab{b}})}\BibitemShut {NoStop}%
\bibitem [{\citenamefont {Takeuchi}\ \emph {et~al.}(2007)\citenamefont
  {Takeuchi}, \citenamefont {Kuroda}, \citenamefont {Chate},\ and\
  \citenamefont {Sano}}]{TKCS07}%
  \BibitemOpen
  \bibfield  {author} {\bibinfo {author} {\bibfnamefont {K.~A.}\ \bibnamefont
  {Takeuchi}}, \bibinfo {author} {\bibfnamefont {M.}~\bibnamefont {Kuroda}},
  \bibinfo {author} {\bibfnamefont {H.}~\bibnamefont {Chate}}, \ and\ \bibinfo
  {author} {\bibfnamefont {M.}~\bibnamefont {Sano}},\ }\href@noop {} {\bibfield
   {journal} {\bibinfo  {journal} {Phys. Rev. Lett.}\ }\textbf {\bibinfo
  {volume} {99}},\ \bibinfo {pages} {234503} (\bibinfo {year}
  {2007})}\BibitemShut {NoStop}%
\bibitem [{\citenamefont {Corte}\ \emph {et~al.}(2008)\citenamefont {Corte},
  \citenamefont {Chaikin}, \citenamefont {Gollub},\ and\ \citenamefont
  {Pine}}]{CCGP08}%
  \BibitemOpen
  \bibfield  {author} {\bibinfo {author} {\bibfnamefont {L.}~\bibnamefont
  {Corte}}, \bibinfo {author} {\bibfnamefont {P.~M.}\ \bibnamefont {Chaikin}},
  \bibinfo {author} {\bibfnamefont {J.~P.}\ \bibnamefont {Gollub}}, \ and\
  \bibinfo {author} {\bibfnamefont {D.~J.}\ \bibnamefont {Pine}},\ }\href@noop
  {} {\bibfield  {journal} {\bibinfo  {journal} {Nature Physics}\ }\textbf
  {\bibinfo {volume} {4}},\ \bibinfo {pages} {420} (\bibinfo {year}
  {2008})}\BibitemShut {NoStop}%
\bibitem [{\citenamefont {Franceschini}\ \emph {et~al.}(2011)\citenamefont
  {Franceschini}, \citenamefont {Filippidi}, \citenamefont {Guazzelli},\ and\
  \citenamefont {Pine}}]{FFGP11}%
  \BibitemOpen
  \bibfield  {author} {\bibinfo {author} {\bibfnamefont {A.}~\bibnamefont
  {Franceschini}}, \bibinfo {author} {\bibfnamefont {E.}~\bibnamefont
  {Filippidi}}, \bibinfo {author} {\bibfnamefont {E.}~\bibnamefont
  {Guazzelli}}, \ and\ \bibinfo {author} {\bibfnamefont {D.~J.}\ \bibnamefont
  {Pine}},\ }\href {\doibase 10.1103/PhysRevLett.107.250603} {\bibfield
  {journal} {\bibinfo  {journal} {Phys. Rev. Lett.}\ }\textbf {\bibinfo
  {volume} {107}},\ \bibinfo {pages} {250603} (\bibinfo {year}
  {2011})}\BibitemShut {NoStop}%
\bibitem [{\citenamefont {Okuma}\ \emph {et~al.}(2011)\citenamefont {Okuma},
  \citenamefont {Tsugawa},\ and\ \citenamefont
  {Motohashi}}]{OkumaTsugawaMotohashi11}%
  \BibitemOpen
  \bibfield  {author} {\bibinfo {author} {\bibfnamefont {S.}~\bibnamefont
  {Okuma}}, \bibinfo {author} {\bibfnamefont {Y.}~\bibnamefont {Tsugawa}}, \
  and\ \bibinfo {author} {\bibfnamefont {A.}~\bibnamefont {Motohashi}},\ }\href
  {\doibase 10.1103/PhysRevB.83.012503} {\bibfield  {journal} {\bibinfo
  {journal} {Phys. Rev. B}\ }\textbf {\bibinfo {volume} {83}},\ \bibinfo
  {pages} {012503} (\bibinfo {year} {2011})}\BibitemShut {NoStop}%
\bibitem [{\citenamefont {Korolev}\ and\ \citenamefont
  {Nelson}(2011)}]{KorolevNelson11}%
  \BibitemOpen
  \bibfield  {author} {\bibinfo {author} {\bibfnamefont {K.~S.}\ \bibnamefont
  {Korolev}}\ and\ \bibinfo {author} {\bibfnamefont {D.~R.}\ \bibnamefont
  {Nelson}},\ }\href {\doibase 10.1103/PhysRevLett.107.088103} {\bibfield
  {journal} {\bibinfo  {journal} {Phys. Rev. Lett.}\ }\textbf {\bibinfo
  {volume} {107}},\ \bibinfo {pages} {088103} (\bibinfo {year}
  {2011})}\BibitemShut {NoStop}%
\bibitem [{\citenamefont {Korolev}\ \emph {et~al.}(2011)\citenamefont
  {Korolev}, \citenamefont {Xavier}, \citenamefont {Nelson},\ and\
  \citenamefont {Foster}}]{KXNF11}%
  \BibitemOpen
  \bibfield  {author} {\bibinfo {author} {\bibfnamefont {K.~S.}\ \bibnamefont
  {Korolev}}, \bibinfo {author} {\bibfnamefont {J.~B.}\ \bibnamefont {Xavier}},
  \bibinfo {author} {\bibfnamefont {D.~R.}\ \bibnamefont {Nelson}}, \ and\
  \bibinfo {author} {\bibfnamefont {K.~R.}\ \bibnamefont {Foster}},\
  }\href@noop {} {\bibfield  {journal} {\bibinfo  {journal} {The American
  Naturalist}\ }\textbf {\bibinfo {volume} {178}},\ \bibinfo {pages} {538}
  (\bibinfo {year} {2011})}\BibitemShut {NoStop}%
\bibitem [{\citenamefont {Odor}\ and\ \citenamefont
  {Menyhard}(2006)}]{OdorMenyhard06}%
  \BibitemOpen
  \bibfield  {author} {\bibinfo {author} {\bibfnamefont {G.}~\bibnamefont
  {Odor}}\ and\ \bibinfo {author} {\bibfnamefont {N.}~\bibnamefont
  {Menyhard}},\ }\href@noop {} {\bibfield  {journal} {\bibinfo  {journal}
  {Phys. Rev. E}\ }\textbf {\bibinfo {volume} {73}},\ \bibinfo {pages} {036130}
  (\bibinfo {year} {2006})}\BibitemShut {NoStop}%
\bibitem [{\citenamefont {Menyh\'ard}\ and\ \citenamefont
  {\'Odor}(2007)}]{MenyhardOdor07}%
  \BibitemOpen
  \bibfield  {author} {\bibinfo {author} {\bibfnamefont {N.}~\bibnamefont
  {Menyh\'ard}}\ and\ \bibinfo {author} {\bibfnamefont {G.}~\bibnamefont
  {\'Odor}},\ }\href {\doibase 10.1103/PhysRevE.76.021103} {\bibfield
  {journal} {\bibinfo  {journal} {Phys. Rev. E}\ }\textbf {\bibinfo {volume}
  {76}},\ \bibinfo {pages} {021103} (\bibinfo {year} {2007})}\BibitemShut
  {NoStop}%
\end{thebibliography}%

\end{document}